\documentclass[twocolumn,,superscriptaddress,letter]{revtex4}
\usepackage{amssymb}
\usepackage{color}
\usepackage{graphicx}
\usepackage{dcolumn}
\usepackage{bm}
\usepackage[header,title,page,titletoc]{appendix}
\usepackage{amsmath}
\usepackage{subfigure}
\usepackage{amsfonts}

\providecommand{\U}[1]{\protect \rule{.1in}{.1in}}

\input{tcilatex}
\begin{document}

\title{The Simulation of Non-Abelian Statistics of Majorana Fermions in
Ising Chain with Z2 Symmetry}
\author{Xiao-Ming Zhao}
\affiliation{Department of Physics, Beijing Normal University, Beijing 100875, China}
\author{Jing Yu}
\affiliation{Faculty of Science, Liaoning Shihua University, Fushun, 113001, P. R. China}
\author{Jing He}
\affiliation{Department of Physics, Hebei Normal University, Hebei, 050024, P. R. China}
\author{Qiu-Bo Cheng}
\affiliation{Department of Physics, Beijing Normal University, Beijing 100875, China}
\author{Ying Liang}
\affiliation{Department of Physics, Beijing Normal University, Beijing 100875, China}
\author{Su-Peng Kou}
\thanks{Corresponding author}
\email{spkou@bnu.edu.cn}
\affiliation{Department of Physics, Beijing Normal University, Beijing 100875, China}

\begin{abstract}
In this paper, we numerically study the non-Abelian statistics of the
zero-energy Majorana fermions on the end of Majorana chain and show its
application to quantum computing by mapping it to a spin model with special
symmetry. In particular, by using transverse-field Ising model with Z2
symmetry, we verify the nontrivial non-Abelian statistics of Majorana
fermions. Numerical evidence and comparison in both Majorana-representation
and spin-representation are presented. The degenerate ground states of a
symmetry protected spin chain therefore previde a promising platform for
topological quantum computation.
\end{abstract}

\maketitle

\section{Introduction}

Majorana fermions have recently attracted much attention due to the
potential application in topological quantum computation \cite{ref1}.
Majorana fermions are particles that are their own antiparticles --- in
contrast with the case for Dirac fermions --- and obey non-Abelian statistics%
\cite{ref1.1,ref1.2,ref1.3}. The exotic properties of Majorana fermions have
attracted increasing interest from researchers\cite%
{ref1.4,ref1.5,ref1.6,ref1.7,ref1.9,ref1.10}. Majorana fermions with zero
energy (Majorana zero modes) had been predicted to be induced by vortices in
two-dimensional spinless $p_{x}+ip_{y}$-wave superconductor\cite%
{ref2,ref3,ref4}, or localize at the ends in a one-dimensional
spin-polarized superconductor chain. Another creative proposal is the
interface of $s$-wave superconductors and topological insulators owing to
the proximity effect. On the other hand, the spin chain has been studied in
depth both theoretically and experimentally. It is known that the
transverse-field Ising model with Z2 symmetry is equivalent to the
one-dimensional spin-polarized superconductor model\cite{ref5}.

In this paper, we numerically study the non-Abelian statistics of the
zero-energy Majorana fermions on the end of Majorana chain and show its
application to quantum computing by mapping it to a spin model with special
symmetry. In particular, by using transverse-field Ising model with Z2
symmetry, we verify the nontrivial quantum statistics of Majorana fermions
numerically using a T-junction wire network, where the Majorana fermions can
be braided by tuning local gates. We may also mimic this T-type braiding in
spin representation numerically, where the two zero-energy Majorana fermion
states correspond to two degenerate ground states of spin chain. In this
way, we provide an easy way of detecting the fundamental non-Abelian
statistics of Majorana fermions, which is useful to quantum computation.

\section{Majorana zero modes in one-dimensional quantum spin model with Z2
symmetry}

It has been recognized that a one-dimensional quantum spin model with Z2
symmetry is equivalent to a one-dimensional superconductor via Jordan-Wigner
transformation. Therefore, we can describe a one-dimensional spin chain
using either spin representation ($\sigma $-representation) or fermion
representation ($\gamma $-representation). Thus, the Majorana fermion and
its statistic property can be represented in either representation.

Here, we start from the one-dimensional Ising chain with Z2 symmetry. The
Hamiltonian of the Ising spin chain is given by%
\begin{equation}
\hat{H}=\sum_{n=1}^{N-1}J_{n,n+1}\sigma _{n}^{x}\sigma
_{n+1}^{x}-\sum_{n=1}^{N}\mu _{n}\vec{\sigma}_{n}
\end{equation}%
where $J_{n,n+1}$ ( $J_{n,n+1}<0$ ) is the Ising coupling constant between
two nearest-neighbour (NN) sites $n,$ $n+1$, $\mu _{n}$ ( $\mu _{n}>0$ ) is
the strength of external field on site $n$, and $N$ is the total lattice
number of the Ising chain. We then introduce the spin operators $\sigma
_{n}^{+},$ $\sigma _{n}^{-}$, $\sigma _{n}^{x}=\sigma _{n}^{+}+\sigma
_{n}^{-},$ and $\sigma _{n}^{y}=i(\sigma _{n}^{+}-\sigma _{n}^{-})$. The Z2
symmetry is charaterized by as spin rotation symmetry $\hat{R}=e^{i\pi
\sum_{n=1}^{N}\sigma _{n}^{z}}$, i.e.,
\begin{equation*}
\hat{R}\hat{H}\hat{R}^{-1}=\hat{H}.
\end{equation*}
Thus, to guarantee the Z2 symmetry, the external field should be along $z$%
-direction, or $\vec{\sigma}_{n}\rightarrow \sigma _{n}^{z}$.

\begin{figure}[tph]
\scalebox{0.33}{\includegraphics*[0.1in,1.5in][10.7in,7.1in]{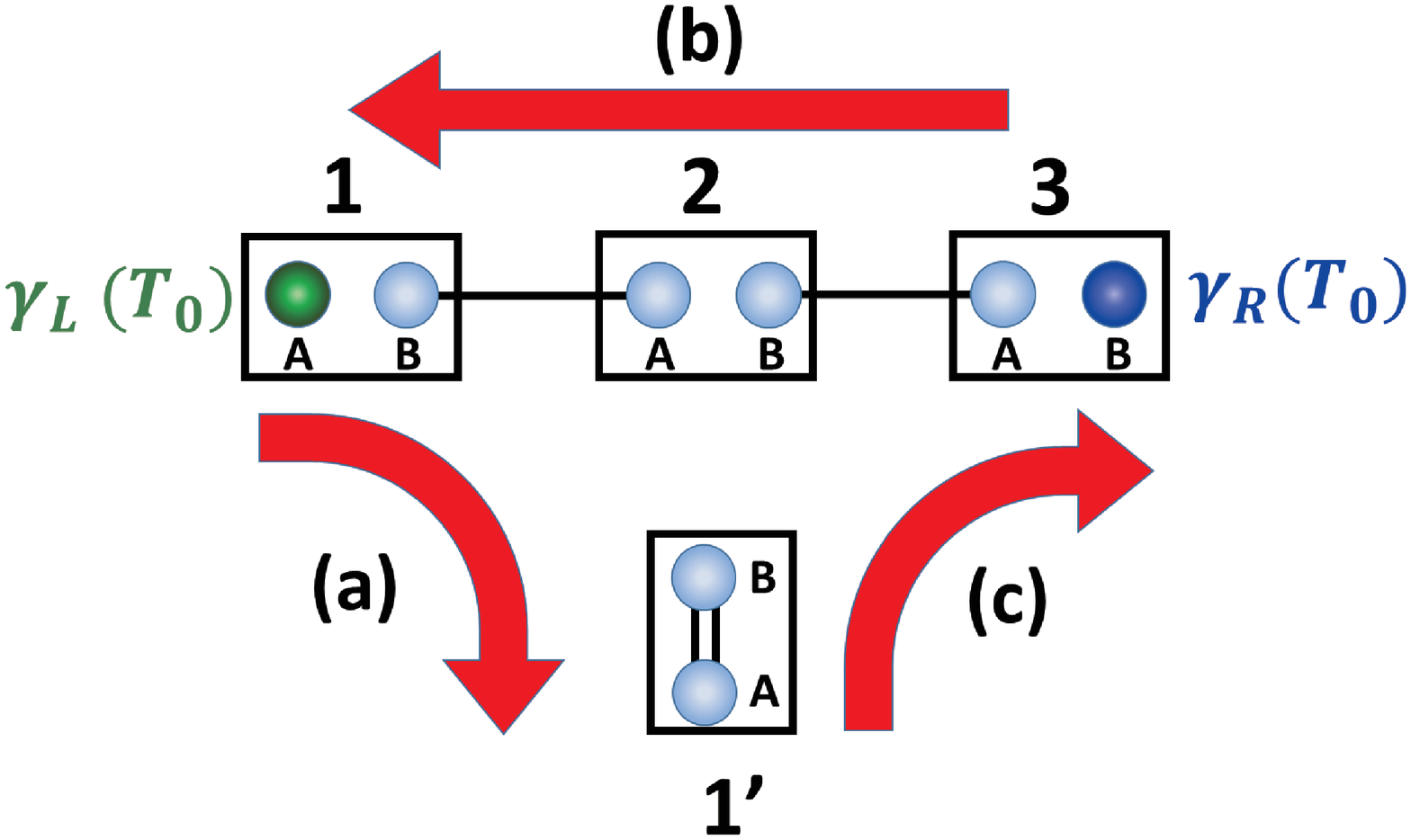}}
\caption{(Color online)\textbf{\ Braiding two Majorana fermions. \ }The
braiding process of Majorana fermions is denoted by thick red arrows. For a
T-junction, there is one spin in each box $1,2,3,$ and $1^{\prime }$. At the
beginning of the braiding two unpaired Majorana fermions locate at the left
end $\protect\gamma _{L}(T_{0})$ (green ball) and right end $\protect\gamma %
_{R}(T_{0})$ (blue ball) of the Majorana chain. We can adiabatically turn on
$\protect\mu _{1}$ $(0\rightarrow \protect\mu _{0})$ and turn off $J_{1B,2A}$
$(J_{0}\rightarrow 0)$ drives the left-end Majorana zero mode from site $1A$
to site $2A$. As the same, we drive the two fermions as follow: \textbf{(a)}
During $T_{0}$ to $T_{2}$ the most left fermion is driven to the bottom site
of box $1^{\prime }$; \textbf{(b)}During $T_{3}$ to $T_{5}$ the most right
fermion is driven to the most left site; \textbf{(c)} During $T_{6}$ to $%
T_{7}$ the bottom one is driven to the most right site. Now the final
Majorana modes are denoted by $\protect\gamma _{L}(T_{7})$, $\protect\gamma %
_{R}(T_{7})$. }
\end{figure}

The Jordan-Wigner transformation is described by\cite{ref5,ref6}%
\begin{align}
\sigma _{n}^{+}& =a_{n}^{+}{\displaystyle\prod\limits_{m=1}^{n-1}}%
a_{l}^{+}a_{l};  \notag \\
\sigma _{n}^{-}& =a_{n}{\displaystyle\prod\limits_{m=1}^{n-1}}a_{l}^{+}a_{l};
\\
\sigma _{n}^{z}& =2a_{n}^{+}a_{n}-1,  \notag
\end{align}%
where $a_{n}^{+},$ $a_{n}$ denote the creation and annihilation operators of
Dirac fermions and obey the anticommutation relation $\left\{
a_{m},a_{n}^{+}\right\} =\delta _{m,n}$. By using Jordan-Wigner
transformation, the Hamiltonian $\hat{H}$ can be written as
\begin{equation}
\hat{H}=\sum_{n=1}^{N-1}J_{n,n+1}(a_{n}-a_{n}^{+})(a_{n+1}+a_{n+1}^{+})-%
\sum_{n=1}^{N}\mu _{n}(2a_{n}^{+}a_{n}-1).
\end{equation}%
Then, the Majorana fermion is defined as
\begin{equation}
\gamma _{n}^{A}=a_{n}^{+}+a_{n},\text{ }\gamma _{n}^{B}=i(a_{n}^{+}-a_{n}),
\end{equation}%
with $\gamma _{n}^{\dag }=\gamma _{n},$ $\left\{ \gamma _{n}^{l},\gamma
_{m}^{l^{\prime }}\right\} =2\delta _{m,n}\delta _{l,l^{\prime }}$. From the
definition, one can see that Majorana fermions are their own antiparticle
and constitute \textquotedblleft half\textquotedblright\ of an ordinary
fermion. We obtain the Hamiltonian in the $\gamma $-representation \cite%
{ref20},%
\begin{equation}
\hat{H}=-i\sum_{n=1}^{N-1}J_{n,n+1}\gamma _{n}^{B}\gamma
_{n+1}^{A}-i\sum_{n=1}^{N}\mu _{n}\gamma _{n}^{A}\gamma _{n}^{B}.
\end{equation}

\begin{figure}[tph]
\scalebox{0.30}{\includegraphics*[0.1in,0.1in][10.7in,7.9in]{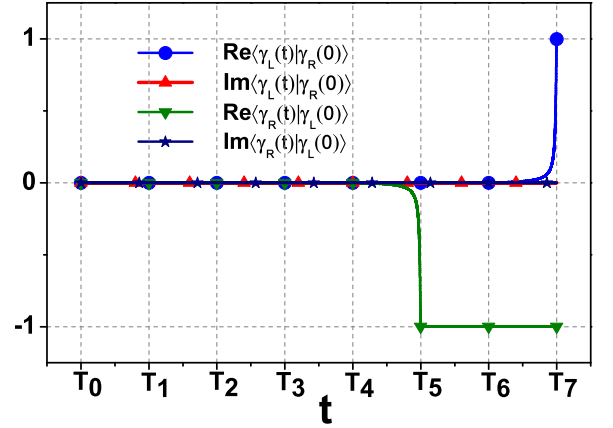}}
\caption{\textbf{Evolution of particle distribution of Majorana modes during
the braiding:} $\left\vert \protect\gamma _{L}(t)\right\rangle $ ($%
\left\vert \protect\gamma _{R}(t)\right\rangle $) is the wave function of
Majorana zero modes during the braiding at time $t$, which localizes in left
(right) end of horizontal chain at time $t=T_{0}=0$. To guarantee the
adiabatic condition, i.e. the process is very slow, we set $%
T_{7}-T_{0}=10000\hbar /J,$ $N_{0}=10^{6}$. After the braiding process, we
have $\left\langle \protect\gamma _{L}(T_{7})|\protect\gamma %
_{L}(0)\right\rangle =\left\langle \protect\gamma _{R}(T_{7})|\protect\gamma %
_{R}(0)\right\rangle =0,$ $\left\vert \protect\gamma _{L}(T_{7})\right%
\rangle =\left\vert \protect\gamma _{R}(0)\right\rangle ,$ $\left\vert
\protect\gamma _{R}(T_{7})\right\rangle =-\left\vert \protect\gamma %
_{L}(0)\right\rangle .$ That is $\protect\gamma _{1}^{A}$ $\rightarrow $ $%
\protect\gamma _{3}^{B}$ and $\protect\gamma _{3}^{B}$ $\rightarrow $ $-%
\protect\gamma _{1}^{A}.$}
\end{figure}

In the fermion representation for the Hamiltonian, we see that two Majorana
fermions on one site $n$ are coupled and the coupling constant is $\mu _{n}$
(e.g., the double dark line links site A and site B inner the box $1^{\prime
}$ in Fig.1) and the two Majorana fermions on the NN sites are linked by $%
J_{n,n+1}$ (e.g., the single dark line between boxes 1, 2 and 3 in Fig.1).
When we adiabatically turn off $\mu _{n}$ at all sites such that its value
decreases from a certain value $\mu _{0}$ to zero ($\mu _{0}\rightarrow 0$),
the Majorana fermions of the chain are only coupled by $J_{n,n+1}$ terms
except for the two Majorana fermions at the ends (e.g., the green and blue
balls in Fig.1).

To characterize the quantum states of Majorana fermions, we introduce the
creation and annihilation operators of Dirac fermions, $d_{n}=(\gamma
_{n+1}^{A}+i\gamma _{n}^{B})/2$, $d_{n}^{\dag }=(\gamma _{n+1}^{A}-i\gamma
_{n}^{B})/2$. The operators of Dirac fermions are combined by two Majorana
fermions at NN sites, i.e., $n$ and $n+1$. Thus, the Majorana fermions at
the left (right) end of the chain $\gamma _{1}^{A}$ $(\gamma _{N}^{B})$
remain unpaired and have zero energy\cite{ref8,ref9}. Here, we focus on the
edge fermion and have
\begin{equation}
d_{end}=\frac{1}{2}(\gamma _{1}^{A}+i\gamma _{N}^{B}),\text{ }d_{end}^{\dag
}=\frac{1}{2}(\gamma _{1}^{A}-i\gamma _{N}^{B}).
\end{equation}%
It is obvious that the edge fermion has zero energy. We now define $%
\left\vert F\right\rangle $ to be a many-body quantum state with occupied
single particle states for $E<0$ and empty single particle states $E\geq 0$.
We therefore introduce a Majorana qubit that consists of two basis states $%
\left\vert 0\right\rangle ,$ $\left\vert 1\right\rangle $ defined as\cite%
{ref6.1}%
\begin{equation}
\left\vert 0\right\rangle \equiv d_{end}\left\vert F\right\rangle ,\text{ }%
\left\vert 1\right\rangle \equiv d_{end}^{\dag }\left\vert 0\right\rangle .
\end{equation}

\section{Numerical verifying non-Abelian statistics of Majorana fermions in $%
\protect\gamma $-representation}

In this part, we numerically study the quantum statistic of the Majorana
fermions located at the end of spin chain by using one-dimensional quantum
Ising model with Z2 symmetry. To explore the quantum statistic of the
Majorana fermions, we take a 4-spin (i.e., 8-$\gamma $) system as an example
and braid the Majorana fermions $\gamma _{1}^{A},$ $\gamma _{3}^{B}$ by
seven steps using the T-type structure (see the illustration in Fig.1),
which is similar to the semiconducting wire networks in Ref.\cite{ref11}.

The parameters $J_{n,n+1}$ and $\mu _{n}$ in the original Hamiltonian are
given by $J_{nB,(n+1)A}=J_{0}$ and $\mu _{n}=\mu _{0}$, respectively. We
first choose the initial state with $J_{nB,(n+1)A}=0$ and $\mu _{n}=0$.
Thus, there must exit two unpaired Majorana zero modes located at the left
end $\left\vert \gamma _{L}(T_{0})\right\rangle =\left\vert \gamma
_{1}^{A}\right\rangle $ (green ball) and right end $\left\vert \gamma
_{R}(T_{0})\right\rangle =\left\vert \gamma _{3}^{B}\right\rangle $ (blue
ball) of the Majorana chain. Here $T_{0}=0$ represents the initial time and $%
T_{n}$ for $n$-th step of braiding process. We denote the quantum states of
Majorana modes by $\left\vert \gamma _{L}(T_{n})\right\rangle ,$ $\left\vert
\gamma _{R}(T_{n})\right\rangle ,$ $n\in (1,7)$. The Hamiltonian of the
system at $T_{0}$ is given by
\begin{equation}
H_{\gamma ,T_{0}}=-iJ_{0}\gamma _{1}^{B}\gamma _{2}^{A}-iJ_{0}\gamma
_{2}^{B}\gamma _{3}^{A}-i\mu _{0}\gamma _{1^{\prime }}^{A}\gamma _{1^{\prime
}}^{B}.
\end{equation}

We then do the braiding process step by step (see Fig.1): \textbf{(a)} $\mu
_{1}|_{_{^{T_{0}}}}^{_{_{_{T_{1}}}}}(0\rightarrow \mu _{0})$, $%
J_{1B,2A}|_{_{^{T_{0}}}}^{_{_{_{T_{1}}}}}(J_{0}\rightarrow 0)$ (This means
we adiabatically turn on $\mu _{1}$ and turn off $J_{1B,2A}$ simultaneously
during the time period $t\in (T_{0},$ $T_{1})$), then $\mu _{1^{\prime
}}|_{_{^{T_{1}}}}^{_{_{_{T_{2}}}}}(\mu _{0}\rightarrow 0)$, $J_{1^{\prime
}B,2A}|_{_{^{T_{1}}}}^{_{_{_{T_{2}}}}}(0\rightarrow J_{0})$. The order of
this braiding process is $1A\rightarrow 2A\rightarrow 1^{\prime }A;$ \textbf{%
(b)} $\mu _{3}|_{_{^{T_{2}}}}^{_{_{_{T_{3}}}}}(0\rightarrow \mu _{0})$, $%
J_{2B,3A}|_{_{^{T_{2}}}}^{_{_{_{T_{3}}}}}(J_{0}\rightarrow 0)$, then $%
J_{1^{\prime }B,2A}|_{_{^{T_{3}}}}^{_{_{_{T_{4}}}}}(J_{0}\rightarrow 0)$, $%
J_{1^{\prime }B,2B}|_{_{^{T_{3}}}}^{_{_{_{T_{4}}}}}(0\rightarrow J_{0})$,
next, $\mu _{1}|_{_{^{T_{4}}}}^{_{_{_{T_{5}}}}}(\mu _{0}\rightarrow 0)$, $%
J_{1B,2A}|_{_{^{T_{4}}}}^{_{_{_{T_{5}}}}}(0\rightarrow J_{0})$. The braiding
order is $3B\rightarrow 2B\rightarrow 2A\rightarrow 1A$; \textbf{(c)} $\mu
_{1^{\prime }}|_{_{^{T_{5}}}}^{_{_{_{T_{6}}}}}(0\rightarrow \mu _{0})$, $%
J_{1^{\prime }B,2B}|_{_{^{T_{5}}}}^{_{_{_{T_{6}}}}}(J_{0}\rightarrow 0)$,
then $\mu _{3}|_{_{^{T_{6}}}}^{_{_{_{T_{7}}}}}(\mu _{0}\rightarrow 0)$, $%
J_{2B,3A}|_{_{^{T_{6}}}}^{_{_{_{T_{7}}}}}(0\rightarrow J_{0})$. The braiding
order is $1^{\prime }A\rightarrow 2B\rightarrow 3B$.

In particular, during the time period $t\in (T_{3},$ $T_{4})$, we have
\begin{align}
H_{\gamma ,T_{3}}& =-iJ_{0}\gamma _{1^{\prime }}^{B}\gamma _{2}^{A}-i\mu
_{0}\gamma _{1}^{A}\gamma _{1}^{B}-i\mu _{0}\gamma _{3}^{A}\gamma _{3}^{B},
\\
H_{\gamma ,T_{4}}& =-iJ_{0}\gamma _{1^{\prime }}^{B}\gamma _{2}^{B}-i\mu
_{0}\gamma _{1}^{A}\gamma _{1}^{B}-i\mu _{0}\gamma _{3}^{A}\gamma _{3}^{B}.
\end{align}%
The operation during this period shifts the Majorana mode from site $2B$ to $%
2A$ which are on the same box $2$.
\begin{figure}[tph]
\scalebox{0.30}{\includegraphics*[0.1in,1.1in][10.7in,7.9in]{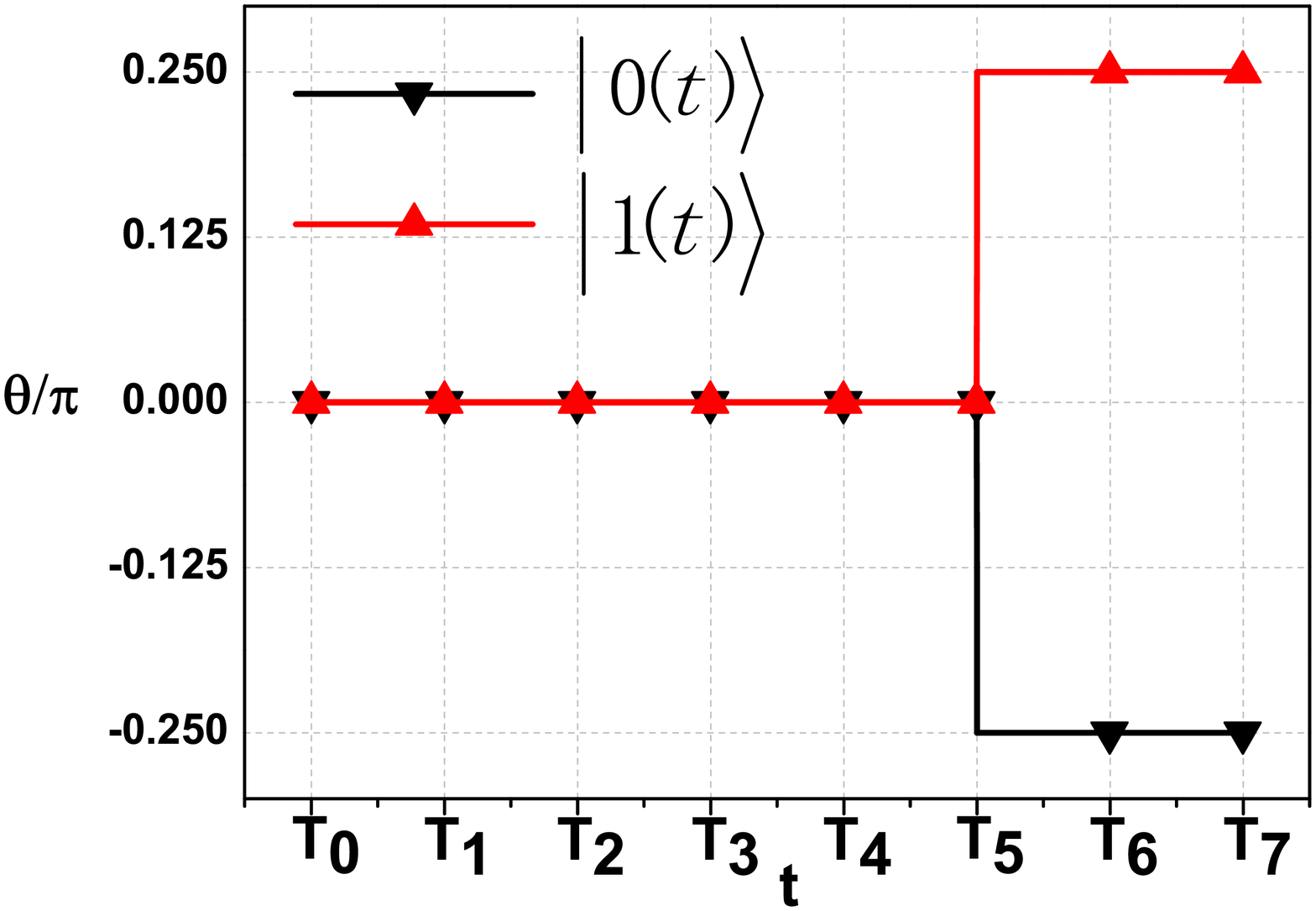}}
\caption{\textbf{Berry phase during the braiding process:} $\left\vert
0(t)\right\rangle $ ($\left\vert 1(t)\right\rangle $) represents one of the
two degenerate ground states in spin system, in which spin direction is
along $\mathbf{x}$-direction (-$\mathbf{x}$) at the time $t=T_{0}=0$ except
for the spin-$1^{\prime }$. The Berry phase of the ground state sharply
changes at $t=T_{5}$ where spin-$1$ and spin-$2$ switch from to $\mathbf{x}$%
-direction to $\mathbf{y}$-direction. As a result, Berry phase changes $%
\frac{\protect\pi }{4}$ during the braiding process. }
\end{figure}

It is well known that the braiding of Majorana modes changes not only the
amplitude but also the phase of the modes. We next focus on the phase
difference of $\left\vert \gamma _{L}(T_{n})\right\rangle ,$ $\left\vert
\gamma _{R}(T_{n})\right\rangle $ before and after the adiabatic braiding
process numerically. We diagonalize the initial Hamiltonian $H_{\gamma
,T_{0}}$ in the $\gamma $-representation and obtain two zero energy modes $%
\left\vert \gamma _{L}(T_{0})\right\rangle =\left\vert \gamma
_{1}^{A}\right\rangle $ and $\left\vert \gamma _{R}(T_{0})\right\rangle
=\left\vert \gamma _{3}^{B}\right\rangle $. We then define a time-evolution
operator
\begin{equation}
U(t)=\hat{T}\left\{ \exp [-i\int_{0}^{t}H(t^{\prime })dt^{\prime }]\right\} ,
\end{equation}%
where $\hat{T}$ is the time ordering operator. Therefore, at the end of the
evolution, we have
\begin{align}
\left\vert \gamma _{L}(T_{7})\right\rangle =& U(T_{7})\left\vert \gamma
_{L}(T_{0})\right\rangle  \notag \\
\text{ \ }\left\vert \gamma _{R}(T_{7})\right\rangle =& U(T_{7})\left\vert
\gamma _{R}(T_{0})\right\rangle .
\end{align}%
To realize the time-evolution numerically, one may discretize the
time-evolution operator employing the times slicing procedure
\begin{equation}
U(T_{7})\approx \hat{T}\bigskip {\displaystyle\prod\limits_{i=0}^{N_{0}}}%
\exp [-iH(t_{i})\triangle t],\text{ \ }\triangle t=\frac{T_{7}-T_{0}}{N_{0}},
\end{equation}%
with $\triangle t\ll \hbar /J,$ and $T_{7}-T_{0}$ being sufficiently large.
We point out that it is crucial to retain the unitarity throughout the
calculation
\begin{equation}
\exp [-iH(t_{i})\triangle t]=A\exp (-i\Lambda \triangle t)A^{\dag },
\end{equation}%
where $H(t_{i})=A\Lambda A^{\dag },$ $A$ is a unitary matrix $AA^{\dag }=I$
and $\Lambda $ is a diagonal matrix. Fig.2 shows the change in $\left\vert
\gamma _{L}(t)\right\rangle ,$ $\left\vert \gamma _{R}(t)\right\rangle $
during the braiding process. It is clearly that $\left\vert \gamma
_{L}(T_{7})\right\rangle =\left\vert \gamma _{R}(0)\right\rangle ,$ $%
\left\vert \gamma _{R}(T_{7})\right\rangle =-\left\vert \gamma
_{L}(0)\right\rangle $. The braiding operation therefore transforms $\gamma
_{1}^{A}$ to $\gamma _{3}^{B}$ and $\gamma _{3}^{B}$ to $-\gamma _{1}^{A}$.

\section{Numerical verifying non-Abelian statistics of Majorana fermions in $%
\protect\sigma $-representation}

In last section, we have verified the non-Abelian statistics numerically in $%
\gamma $-representation and construct a phase gate based on the qubits that
is simple and easily understood\cite{ref11}. We then map the braiding in $%
\gamma $-representation to that in $\sigma $-representation by employing the
Jordan-Wigner transformation \cite{ref13},
\begin{figure}[tph]
\scalebox{0.30}{\includegraphics*[0.1in,1.2in][10.7in,6.9in]{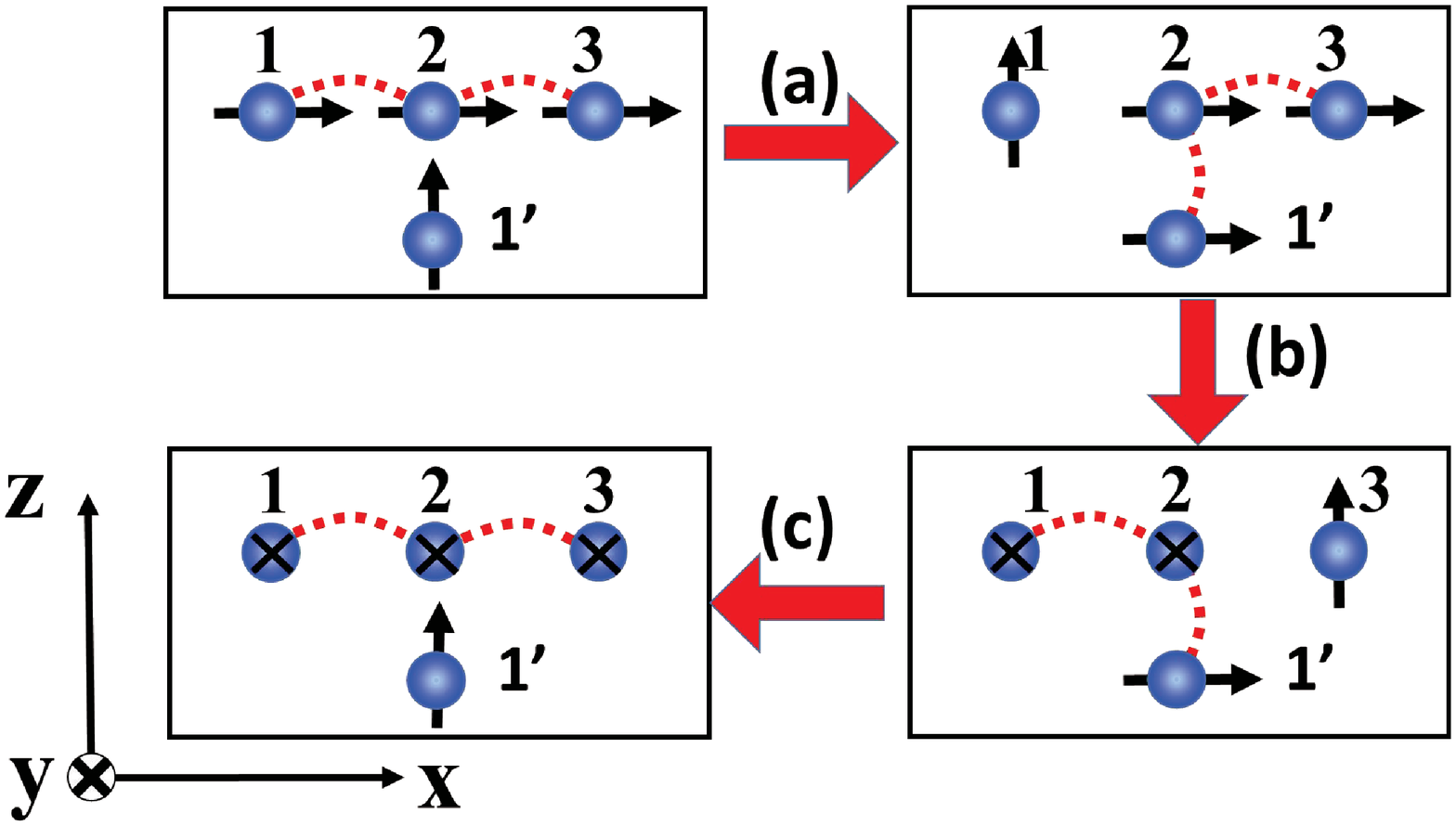}}
\caption{\textbf{The illustration of braiding Majorana fermions in spin
representation: }The first picture is an illustration of one of the two
degenerate ground states in the spin system which are equal to the two
Majorana zero modes in the $\protect\gamma $-representation. Tuning the
parameters using the same processes as in Fig.1\textbf{(a)}-\textbf{(c)}. }
\end{figure}
\begin{equation}
\gamma _{n}^{A}=({\displaystyle\prod\limits_{m=1}^{n-1}}\sigma
_{m}^{z})\sigma _{n}^{x},\text{ \ \ }\gamma _{n}^{B}=i({\displaystyle%
\prod\limits_{m=1}^{n}}\sigma _{m}^{z})\sigma _{n}^{x}.
\end{equation}%
It is obvious that the Majorana fermions $\gamma _{n}^{A}$ and $\gamma
_{n}^{B}$ are non-local in the $\sigma $-representation. When $J_{0}<0$ the
state $\left\vert F\right\rangle $ can be written as%
\begin{equation}
\left\vert F\right\rangle =\left\vert \rightarrow \rightarrow \rightarrow
\right\rangle ,
\end{equation}%
where
\begin{equation}
\left\vert \rightarrow \right\rangle =\frac{\sqrt{2}}{2}\left(
\begin{array}{c}
1 \\
1%
\end{array}%
\right) ,\text{ \ }\left\vert \leftarrow \right\rangle =\frac{\sqrt{2}}{2}%
\left(
\begin{array}{c}
1 \\
-1%
\end{array}%
\right) .
\end{equation}%
Then the two basis states $\left\vert 0\right\rangle $, $\left\vert
1\right\rangle $ of Majorana qubit are represented in $\sigma $%
-representation as
\begin{align}
\left\vert 0\right\rangle & =\frac{1}{2}(\gamma _{1}^{A}+i\gamma
_{3}^{B})\left\vert F\right\rangle  \notag \\
& =\frac{1}{2}(\sigma _{1}^{x}+i(i{\displaystyle\prod\limits_{m=1}^{3}}%
\sigma _{m}^{z})\sigma _{3}^{x})\left\vert F\right\rangle \\
& =\frac{\sqrt{2}}{2}(\left\vert \rightarrow \rightarrow \rightarrow
\right\rangle -\left\vert \leftarrow \leftarrow \leftarrow \right\rangle ),
\notag
\end{align}%
\begin{align}
\left\vert 1\right\rangle & =\frac{1}{2}(\gamma _{1}^{A}-i\gamma
_{N}^{B})\left\vert 0\right\rangle  \notag \\
& =\frac{1}{2}(\sigma _{1}^{x}-i(i{\displaystyle\prod\limits_{m=1}^{3}}%
\sigma _{m}^{z})\sigma _{3}^{x})\left\vert 0\right\rangle \\
& =\frac{\sqrt{2}}{2}(\left\vert \rightarrow \rightarrow \rightarrow
\right\rangle +\left\vert \leftarrow \leftarrow \leftarrow \right\rangle ).
\notag
\end{align}%
Thus, the two quantum states of Majorana fermions correspond to two
degenerate ground states of 1D transverse Ising model with Z2 symmetry.
\begin{table}[tbph]
\begin{tabular}{|m{0.5in}|c|c|}
\hline
& $\gamma$-representation & $\sigma$-representation \\ \hline
Fermion operator & $\gamma_{n}^{A},\gamma_{n}^{B}$ & $(\prod
\limits_{m=1}^{n-1}\sigma_{m}^{z})\sigma_{n}^{x},i(\prod
\limits_{m=1}^{n}\sigma_{m}^{z})\sigma_{n}^{x}$ \\ \hline
String operator & $i\prod \limits_{m=1}^{N}\gamma_{m}^{A}\gamma_{m}^{B}$ & $%
\prod \limits_{m=1}^{N}\sigma_{m}^{z}$ \\ \hline
Basis & $\frac{1}{2}(\gamma_{1}^{A}+i\gamma_{N}^{B})\left \vert F\right
\rangle $ & $\frac{\sqrt{2}}{2}(\left \vert \rightarrow \rightarrow \cdot
\rightarrow \right \rangle -\left \vert \leftarrow \leftarrow \cdot
\leftarrow \right \rangle )$ \\
state & $\frac{1}{2}(\gamma_{1}^{A}-i\gamma_{N}^{B})\left \vert 0\right
\rangle $ & $\frac{\sqrt{2}}{2}(\left \vert \rightarrow \rightarrow \cdot
\rightarrow \right \rangle +\left \vert \leftarrow \leftarrow \cdot
\leftarrow \right \rangle )$ \\ \hline
Braiding & Majorana modes & Spin rotation $\pi/2$ \\
process & exchange & \ around z-axis \\ \hline
Braiding & $\gamma_{1}^{A}\rightarrow \gamma_{N}^{B}$ & $\left \vert 0\right
\rangle _{\sigma}\rightarrow e^{i\pi/2}\left \vert 0\right \rangle _{\sigma}$
\\
results & $\gamma_{N}^{B}\rightarrow-\gamma_{1}^{A}$ & $\left \vert 1\right
\rangle _{\sigma}\rightarrow \left \vert 1\right \rangle _{\sigma}$ \\ \hline
\end{tabular}%
\caption{\textbf{The comparison of braiding operations of two Majorana
fermions of 1D transverse Ising model with Z2 symmetry in }$\protect\gamma $%
\textbf{-representation and that in }$\protect\sigma $\textbf{%
-representation. }}
\end{table}

Analogy to the previous braiding process, we can obtain the Hamiltonian of
the 4-spin system in different time periods $T_{n}$ as
\begin{align}
H_{\sigma ,T_{0}}& =J_{0}\sigma _{1}^{x}\sigma _{2}^{x}-J_{0}\sigma
_{2}^{x}\sigma _{3}^{x}-\mu _{0}\sigma _{1^{\prime }}^{z},  \notag \\
H_{\sigma ,T_{3}}& =J_{0}\sigma _{1^{\prime }}^{x}\sigma _{2}^{x}-\mu
_{0}\sigma _{1}^{z}-\mu _{0}\sigma _{3}^{z}, \\
H_{\sigma ,T_{4}}& =J_{0}\sigma _{1^{\prime }}^{x}\sigma _{2}^{y}-\mu
_{0}\sigma _{1}^{z}-\mu _{0}\sigma _{3}^{z}.  \notag
\end{align}%
For a state $\left\vert \psi (\tau )\right\rangle $, we can define the Berry
phase \cite{ref10} as%
\begin{equation}
\theta =i\int_{T_{0}}^{T_{7}}\left\langle \psi (\tau )\right\vert \frac{d}{%
d\tau }\left\vert \psi (\tau )\right\rangle d\tau .
\end{equation}%
The changes of $\theta $ for $\left\vert 0\right\rangle $, $\left\vert
1\right\rangle $ in the process of evolution are shown in Fig.3. We find
that $\theta _{0}=\frac{\pi }{4}$, $\theta _{1}=-\frac{\pi }{4}$, i.e. the
phase difference of $\left\vert 0(t)\right\rangle $, $\left\vert
1(t)\right\rangle $ is $\frac{\pi }{2},$ so we have
\begin{equation}
\binom{\left\vert 0\right\rangle }{\left\vert 1\right\rangle }\rightarrow
\left(
\begin{array}{cc}
e^{i\frac{\pi }{2}} & 0 \\
0 & 1%
\end{array}%
\right) \binom{\left\vert 0\right\rangle }{\left\vert 1\right\rangle }.
\end{equation}%
The braiding process equals to rotating the spin at site $1,$ $2,$ $3$ in
the $x$-$y$ plane from $x$-direction to $y$-direction. This process is shown
in Fig.4, in which the three processes correspond to that of \textbf{(a)},
\textbf{(b)}, \textbf{(c)} in Fig1. We can also describe the results of the
evolution as follow
\begin{align}
& \left\vert \rightarrow \rightarrow \rightarrow \right\rangle  \notag \\
\rightarrow & \frac{\sqrt{2}}{2}(\left\vert \uparrow \uparrow \uparrow
\right\rangle +e^{i\frac{\pi }{2}}\left\vert \downarrow \downarrow
\downarrow \right\rangle ) \\
=& \frac{\sqrt{2}}{2}(e^{i\frac{\pi }{4}}\left\vert \rightarrow \rightarrow
\rightarrow \right\rangle +e^{-i\frac{\pi }{4}}\left\vert \leftarrow
\leftarrow \leftarrow \right\rangle ),  \notag
\end{align}%
while the other ground state have a similar changes
\begin{align}
& \left\vert \leftarrow \leftarrow \leftarrow \right\rangle  \notag \\
\rightarrow & \frac{\sqrt{2}}{2}(\left\vert \uparrow \uparrow \uparrow
\right\rangle +e^{-i\frac{\pi }{2}}\left\vert \downarrow \downarrow
\downarrow \right\rangle ) \\
=& \frac{\sqrt{2}}{2}(e^{-i\frac{\pi }{4}}\left\vert \rightarrow \rightarrow
\rightarrow \right\rangle +e^{i\frac{\pi }{4}}\left\vert \leftarrow
\leftarrow \leftarrow \right\rangle ).  \notag
\end{align}

Finally, we show a comparison in Tab.1, in which the fermion operator,
string operator, basis state, braiding process and braiding results are
illustrated in both $\gamma $-representation and $\sigma $-representation.
In brief, the braiding of Majorana fermion can be simulated by braiding a
corresponding Ising chain with Z2 symmetry.

\section{Conclusion}

In the end, we draw the conclusion. In this paper, we pointed out that the
transverse-field Ising model with Z2 symmetry may simulate one-dimensional
Majorana chain to braid Majorana fermions. On the one hand, in $\gamma $%
-representation by doing Jordon-Wigner transformation, two zero-energy
Majorana fermions are localized at the left and right ends of the Majorana
chain. We get numerically the transformations $\gamma _{1}^{A}\rightarrow
\gamma _{3}^{B}$ and $\gamma _{3}^{B}$ $\rightarrow $ $-\gamma _{1}^{A}$ by
braiding two Majorana fermions in a T-junction. On the other hand, in $%
\sigma $-representation, the two degenerate ground states correspond to the
degenerate quantum states of two Majorana fermions. The braiding process of
the Majorana zero modes is exactly mapped to switch the spin direction from
the $x$-axis to the $y$-axis in the $x$-$y$ plane. Tab.1 shows the
correspondence between the two representations. Therefore, the Ising chain
with Z2 symmetry can be employed to construct the phase gate in quantum
computation.

\begin{acknowledgments}
This work is supported by National Basic Research Program of China (973
Program) under the grant No. 2011CB921803, 2012CB921704 and NSFC Grant
No.11174035, 11474025, 11404090, 11304136, SRFDP, the Fundamental Research
Funds for the Central Universities, NSF-Hebei Province under Grant No.
A2015205189 and NSF-Hebei Education Department under Grant No. QN2014022.
\end{acknowledgments}

\end{document}